\documentclass[3p,times,twocolumn]{elsarticle}

\usepackage{ecrc}

\volume{00}

\firstpage{1}

\journalname{Nuclear Physics B Proceedings Supplements}

\runauth{O. Yasuda}

\jnltitlelogo{Nuclear Physics B Proceedings Supplements}

\usepackage{amssymb}

\usepackage{graphicx}

\begin{document}

\begin{frontmatter}



\dochead{}

\title{Phenomenology of neutrino oscillation \\
---Brief overview and its relevance to tau physics---}


\author{Osamu Yasuda}

\address{Department of Physics, Tokyo Metropolitan University,
Minami-Osawa, Hachioji, Tokyo 192-0397, Japan}

\begin{abstract}

Status of neutrino oscillation study is briefly reviewed.
Some aspects relevant to tau physics are also described.
\end{abstract}

\begin{keyword}
neutrino oscillation \sep parameter degeneracy \sep
new physics

\end{keyword}

\end{frontmatter}


\section{Status of study of the three flavor neutrino oscillation}

In the standard framework of three massive neutrinos,
the parameters which describe neutrino oscillation
phenomena are three mixing angles
$\theta_{12}$, $\theta_{23}$, $\theta_{13}$,
a CP phase $\delta$, and two mass squared
differences $\Delta m^2_{21}$, $\Delta m^2_{31}$.
The parameters ($|\Delta m_{31}^2|$, $\theta_{23}$)
and ($\Delta m_{21}^2$, $\theta_{12}$) were
determined by the experiments of atmospheric and
accelerator neutrinos, and by those of solar and
long baseline reactor neutrinos,
respectively\,\cite{Beringer:1900zz}.
On the other hand, the value of $\theta_{13}$
was determined recently by the two accelerator neutrino
experiments, T2K\,\cite{Abe:2011sj} and MINOS\,\cite{Adamson:2011qu}
and by the three reactor neutrino
experiments, Double-CHOOZ\,\cite{Abe:2011fz},
Daya Bay\,\cite{An:2012eh} and Reno\,\cite{Ahn:2012nd}.
The results of the global analysis are
given in Refs.\,\cite{Tortola:2012te,Fogli:2012ua,GonzalezGarcia:2012sz}.

The quantities which are not yet determined are
the pattern of mass hierarchy (in other words
the sign of $\Delta m_{31}^2$), the CP phase $\delta$
and the octant of $\theta_{23}$ (in other words
the sign of $\theta_{23}-\pi/4$).\footnote
{It was reported in
Refs.\,\cite{Itow,Tortola:2012te,Fogli:2012ua,GonzalezGarcia:2012sz}
that the atmospheric neutrino data
suggests a slight hint of deviation of
$\theta_{23}$ from $\pi/4$.  However,
we need more statistics before
we draw any conclusion.
}
The next things to do is to determine
the pattern of mass hierarchy and
the octant of $\theta_{23}$, before we achieve
the final goal of the neutrino oscillation study,
i.e., measurement of the CP phase $\delta$.
These tasks are expected to be done 
in the future experiments.
These experiments include so-called super beam
experiments (such as no$\nu$a\,\cite{Ayres:2004js},
T2K phase 2\,\cite{Itow:2001ee,Abe:2011ts,Badertscher:2008bp},
LBNE\,\cite{Goon:2012if},
LBNO\,\cite{LBNO}), in which neutrinos are
produced in pion decays, 
a neutrino factory\,\cite{Geer:1997iz,Bandyopadhyay:2007kx},
in which neutrinos are
produced in muon decays,
or a beta beam\,\cite{Zucchelli:2002sa},
in which neutrinos are
produced in beta decays of radioactive
isotopes.

\section{Paremeter degeneracy of the
neutrino oscillation parameters}
In the accelerator neutrino experiments,
the appearance oscillation probabilities
$P(\nu_\mu \to \nu_e)$ and
$P(\bar{\nu}_\mu \to \bar{\nu}_e)$
($P(\nu_e \to \nu_\mu)$ and
$P(\bar{\nu}_e\to \bar{\nu}_\mu)$
in the case of a neutrino factory and a beta beam)
are measured, and one would naively expect
that the oscillation probabilities of these
two channels are sufficient to determine
the CP phase $\delta$.
It is known, unfortunately, that even if the values of the oscillation
probabilities $P(\nu_\mu \to \nu_e)$ and
$P(\bar{\nu}_\mu \to \bar{\nu}_e)$ are exactly given,
we cannot determine uniquely the values of the
oscillation parameters due to eightfold parameter degeneracy.
This eightfold parameter degeneracy\,\cite{Barger:2001yr}
consists of
three kinds of degeneracies,
the intrinsic degeneracy\,\cite{BurguetCastell:2001ez},
the sign degeneracy\,\cite{Minakata:2001qm}
($\Delta m^2_{31}\leftrightarrow -\Delta m^2_{31}$)
and the octant degeneracy\,\cite{Fogli:1996pv}
($\theta_{23}\leftrightarrow \pi/2-\theta_{23}$).

\begin{figure}[t]
    \includegraphics[width=0.7\linewidth]{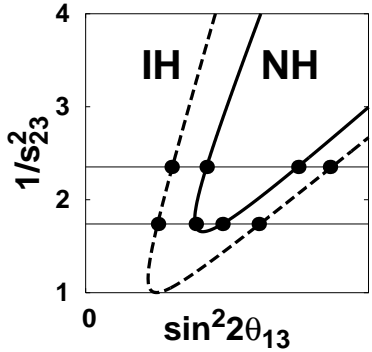}
    \caption{\label{fig:1}\sl%
The eight-fold degeneracy in the $(\sin^22\theta_{13}, 1/s^2_{23})$
plane.  When the two probabilities $P(\nu_\mu \to
\nu_e)=$ const. and $P(\bar{\nu}_\mu \to \bar{\nu}_e)=$ const.
are given, it gives us a quadratic curve for each mass
hierarchy.  NH (solid) and IH (dashed) stand for normal
($\Delta m_{31}^2>0$)
and inverted ($\Delta m_{31}^2<0$) hierarchy, respectively.
}
\end{figure}

To see the eightfold degeneracy, it is convenient for
the plot to give eight different points for different eight solutions.
In Ref.\,\cite{Yasuda:2004gu} it was shown that the solution
specified by $P\equiv P(\nu_\mu \to \nu_e)=$ const. and
$\bar{P}\equiv P(\bar{\nu}_\mu \to \bar{\nu}_e)=$ const. gives
a quadratic curve (a hyperbola in most cases)
in the $(\sin^22\theta_{13}, 1/s^2_{23})$ plane.
In Fig.\ref{fig:1}, $P=$ const. and $\bar{P}=$ const.
gives two different quadratic curves (due to the sign degeneracy)
depending on the pattern of mass hierarchy,
and each curve in general has two intersections
(due to the intrinsic degeneracy) with
one of two horizontal lines
$1/s^2_{23}=2/(1\pm\sqrt{1-\sin^22\theta_{23}})$
(two lines appear due to the octant degeneracy).

\begin{figure}[t]
    \includegraphics[width=1.1\linewidth]{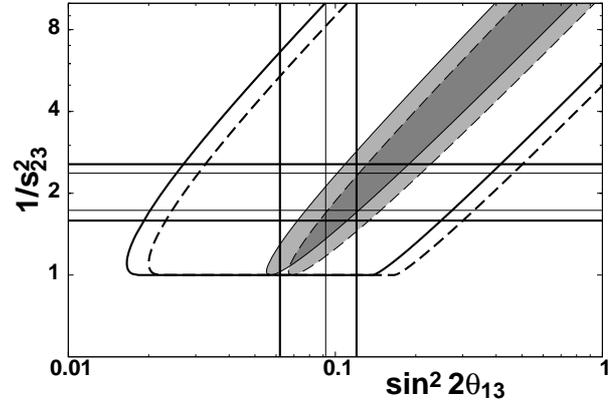}
    \caption{\label{fig:t2k}\sl%
The allowed regions at best-fit (the shaded area
bounded by the thin quadratic curve) and at 90\%CL
(the wider area bounded by the thick quadratic curves and
the line $1/s^2_{23}=1$)
from the T2K appearance result\,\cite{Abe:2011sj}.
The regions bounded by solid (dashed) curves are for normal
(inverted) hierarchy.  The horizontal (vertical) band
$1.6\lesssim1/s^2_{23}\lesssim 2.6$
($0.06\lesssim\sin^22\theta_{13}\lesssim 0.012$) bounded by
thick straight lines stands for the allowed region at 90\%CL
of the atmospheric (reactor) neutrino experiments, respectively.
The horizontal (vertical) thin straight line(s)
$1/s^2_{23}\simeq 1.7$, $1/s^2_{23}\simeq 2.4$
($\sin^22\theta_{13}\simeq 0.09$)
stand for the best-fit
value(s) for the atmospheric (reactor) neutrino data.
}
\end{figure}

\begin{figure}[t]
    \includegraphics[width=1.0\linewidth]{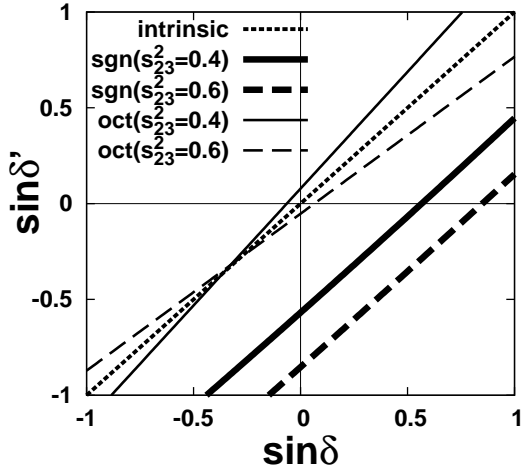}
    \caption{\label{fig:2}\sl%
      The dependence of the value of the CP phase $\delta'$
which is obtained with a wrong ansatz with respect to
the intrinsic, sign and octant degeneracies
as a function of
the true value of $\delta$ from the T2K 
appearance results on $\nu_\mu \to\nu_e$
and $\bar{\nu}_\mu \to \bar{\nu}_e$.
$\sin^22\theta_{13}=0.1$, $\sin^22\theta_{23}=0.96$
is assumed.
}
\end{figure}

Each quadratic curve in Fig.\ref{fig:1} shrinks to a straight line
when the experiment is performed at the oscillation maximum
$|\Delta m^2_{31}|L/4E=\pi/2$,
and this is the case at the T2K experiment.  However,
the present result\,\cite{Abe:2011sj} is for the neutrino
mode $\nu_\mu \to\nu_e$ only, and $P=$ const. gives the inside
of the region bounded by some quadratic curve
(the shaded area bounded by the thin quadratic curve
in Fig.\ref{fig:t2k}), instead of a straight line.
Since the T2K result comes with the experimental error,
the allowed region at 90 \%CL becomes (bounded by the thick quadratic curve and
the line $1/s^2_{23}=1$)
much wider than that by the best-fit value in Fig.\ref{fig:t2k}.

In the future the T2K experiment will measure both
$P$ and $\bar{P}$ with intense
neutrino beams, and the error is expected to be reduced.
Even with the reduced error, however, it is not obvious
whether parameter degeneracy can be resolved.
It turns out that the sign degeneracy is most
serious to determine $\delta$.  This can be seen
from Fig.\ref{fig:2}, where
the value of $\sin\delta'$, which is obtained
with a wrong ansatz with respect to the intrinsic,
sign, and octant degeneracies, respectively,
is plotted as a function
of the true value of $\sin\delta$ in the case of
T2K setup for $\sin^22\theta_{13}=0.1$ 
and $\sin^22\theta_{13}=0.96$.
Since the sign degeneracy can be lifted with
a long baseline experiment with
a baseline length $\gtrsim$ 1000km because of a large
matter effect, it is expected to
be resolved by future experiments with
very long baselines.\footnote{
See Table 1 of Ref.\,\cite{Bertolucci:2012fb}
for a list of the future experiments
which have a potential of resolving the sign
degeneracy.}

\section{New physics and $\nu_\tau$ detection}

It is expected that
the accelerator long baseline experiments with intense
neutrino beams will enable us not only to measure
precisely
the oscillation parameters of the three flavor framework
but also to probe new physics by looking for deviation from
the standard scenario with three massive neutrinos.

In particular, at accelerator neutrino experiments with high
energy neutrinos we can study
the channels $\nu_\mu\to\nu_\tau$ and
$\nu_e\to\nu_\tau$
with OPERA-like detectors.
These experiments include the MINSIS
proposal~\cite{MINSIS,Alonso:2010wu},
in which the NUMI beam is used to look for $\nu_\tau$
at short baseline,
and a neutrino
factory\,\cite{Geer:1997iz,Bandyopadhyay:2007kx}.\footnote{
The current baseline of the neutrino factory~\cite{Choubey:2011zzq}
assumes 10GeV for the muon energy and 2000km for
the baseline length, and detection
of $\nu_\tau$ is expected to be difficult in this scenario.
However, $\nu_\tau$ detection may be possible
in a future extension to the high energy
neutrino factory.}

These channels are expected to give stronger
constraints on new physics than those which
are obtained by the present experiments.
Here I will describe a few examples of new physics,
where $\nu_\tau$ detection can improve on the limits.

\subsection{New Physics at source and detector}

Let us suppose that a non-standard interaction
${\epsilon}_{\alpha\beta} G_F
(\overline{\nu}_\alpha \gamma_\mu \ell_\beta)\,
(\overline{f} \gamma^\mu f')$
exists due to new physics, where
$\epsilon_{\alpha\beta}~(\alpha, \beta=e, \mu, \tau)$ 
is the coefficient of the non-standard interaction, 
normalized in terms of the weak scale,
$\nu_\alpha$ and $\ell_\alpha$ are neutrinos
and charged leptons of flavor $\alpha$, and
$f$ and $f'$ stand for fermions (the only relevant
ones are electrons, u and d quarks).  Then
this interaction predicts the exotic reactions
such as $\pi^+\to\mu^++\nu_e$ or
$\nu_\mu+n\to p+\tau^-$ in the
process of production and detection of
neutrinos\,\cite{Grossman:1995wx}.

Sensitivity to $\epsilon_{\alpha\beta}$ has been
investigated by several groups.  Among others,
Ref.~\cite{Ota:2001pw} showed that 
the sensitivity of a neutrino factory
to ${\epsilon}_{\alpha\tau}$ ($\alpha=e, \mu$)
is excellent.\footnote{
See Ref.\,\cite{Davidson:2003ha} for the bounds
on ${\epsilon}_{\alpha\beta}$ by various experiments.}

\subsection{Violation of unitarity}

It is known~\cite{Antusch:2006vwa} that
in generic see-saw models the kinetic term
gets modified after integrating out
the right handed neutrino and unitarity is expected to be violated.
In the case of the so-called minimal unitarity violation,
in which only three light neutrinos are involved and
sources of unitarity violation are assumed to appear only
in the neutrino sector,
unitarity violation is strongly constrained.
Its constraint mostly comes
from the bounds of rare decays of charged leptons,
and deviation from unitarity
is smaller than ${\cal O}$(1\%)~\cite{Antusch:2006vwa}.

When the mixing matrix $N$ is nonunitary, 
deviation from unitarity is expressed as $NN^\dagger-1$.
Ref.~\cite{Antusch:2006vwa,FernandezMartinez:2007ms,Antusch:2009pm} studied
sensitivity to $(NN^\dagger-1)_{\mu\tau}$ at a neutrino factory,
and the bound is stronger than that of the present one
from $\tau\to\mu\gamma$.

\subsection{Light sterile neutrinos}
The anomaly which was announced
by the LSND group\,\cite{Aguilar:2001ty} 
would imply mass squared
difference of ${\cal O}$(1) eV$^2$
if it is interpreted as a phenomenon due to neutrino oscillation
$\bar{\nu}_\mu\to\bar{\nu}_e$.
The standard three flavor scheme has only two
independent mass squared differences, i.e.,
$\Delta m^2_{21}\simeq 8\times 10^{-5}$eV$^2$
($|\Delta m^2_{31}|\simeq2.4\times 10^{-3}$eV$^2$)
for the solar (atmospheric) neutrino
oscillation.
To accommodate a neutrino oscillation scheme to the LSND
anomaly, therefore, the extra state should be
introduced.  This extra state should be sterile neutrino, which is
singlet with respect to the gauge group of the Standard Model,
because the number of weakly interacting light neutrinos
should be three from the LEP data\,\cite{Beringer:1900zz}.
To test the LSND anomaly, the MiniBooNE experiment
has been performed, but their
results\,\cite{AguilarArevalo:2007it,AguilarArevalo:2010wv}
seem to be inconclusive.  On the other hand, the
flux of the reactor neutrino was recalculated in
Ref.\,\cite{Mueller:2011nm} and it was claimed
that the normalization is shifted by about +3\% on
average.  This claim is qualitatively consistent with
an independent calculation in Ref.\,\cite{Huber:2011wv}.
If their claim on the reactor neutrino flux is correct,
then neutrino oscillation with $\Delta m^2\gtrsim$1eV$^2$
may be concluded from a re-analysis of 19 reactor neutrino
results at short baselines\,\cite{Mention:2011rk}.
This is called reactor anomaly.
Furthermore, it was pointed out in Ref.\,\cite{Giunti:2007xv} that
the measured and predicted $^{71}$Ge production rates
differ in the Gallium radioactive source experiments GALLEX and SAGE,
and this is called Gallium anomaly.
The anomalies of
LSND, reactor and Gallium constitute the main motivation for study
of sterile neutrino oscillations.

Sensitivity to sterile neutrino mixing at the MINSIS
proposal was studied in Ref.\,\cite{Alonso:2010wu},
and it was shown that MINSIS can improve on
the current bound by a factor 100
for the $\nu_\mu\to\nu_\tau$ channel.
In a neutrino scheme with one sterile neutrino state,
there are three CP phases,
and it was shown in Ref.\,\cite{Donini:2008wz} that
the largest CP violation can potentially occur
in the $\nu_\mu\to\nu_\tau$ channel with the present
constraints from other experiments.
Recently a project called nuSTORM\,\cite{Kyberd:2012iz} was proposed
as a low energy prototype for a neutrino factory.
Although the low energy neutrino beam of nuSTORM
does not allow us to create $\tau$'s, it has
a good sensitivity to the sterile neutrino mixing angles.

\section{Summary}
In the standard framework of
three massive neutrinos, all the three
mixing angles have been determined, and
the remaining quantities to be measured
are the sign of $\Delta m_{31}^2$, 
the sign of $\theta_{23}-\pi/4$, and
the CP phase $\delta$.  These quantities
are expected to be determined in the future
experiments, by resolving parameter degeneracies.
If the neutrino energy is sufficiently high
as in the MINSIS proposal or the high energy
option of a neutrino factory,
then one can detect $\nu_\tau$ with an OPERA-like
detector.  In this case, we can probe new physics,
such as non-standard interactions, unitarity violation,
or light sterile neutrino scenarios, so
$\nu_\tau$ detection in neutrino oscillations
deserves further study.

\section*{Acknowledgments}
I would like to thank the organizers for invitation and hospitality
during the workshop.
This research was partly
supported by a Grant-in-Aid for Scientific Research of the
Ministry of Education, Science and Culture, under Grant
No. 24540281.





\begin{thebibliography}{00}

\bibitem{Beringer:1900zz}
  J.~Beringer {\it et al.}  [Particle Data Group Collaboration],
  Phys.\ Rev.\ D {\bf 86} (2012) 010001.

\bibitem{Abe:2011sj}
  K.~Abe {\it et al.}  [T2K Collaboration],
  Phys.\ Rev.\ Lett.\  {\bf 107} (2011) 041801
  [arXiv:1106.2822 [hep-ex]].

\bibitem{Adamson:2011qu}
  P.~Adamson {\it et al.}  [MINOS Collaboration],
  Phys.\ Rev.\ Lett.\  {\bf 107} (2011) 181802
  [arXiv:1108.0015 [hep-ex]].

\bibitem{Abe:2011fz}
  Y.~Abe {\it et al.}  [DOUBLE-CHOOZ Collaboration],
  Phys.\ Rev.\ Lett.\  {\bf 108} (2012) 131801
  [arXiv:1112.6353 [hep-ex]].

\bibitem{An:2012eh}
  F.~P.~An {\it et al.}  [DAYA-BAY Collaboration],
  Phys.\ Rev.\ Lett.\  {\bf 108} (2012) 171803
  [arXiv:1203.1669 [hep-ex]].

\bibitem{Ahn:2012nd}
  J.~K.~Ahn {\it et al.}  [RENO Collaboration],
  Phys.\ Rev.\ Lett.\  {\bf 108} (2012) 191802
  [arXiv:1204.0626 [hep-ex]].

\bibitem{Tortola:2012te}
  D.~V.~Forero, M.~Tortola and J.~W.~F.~Valle,
  Phys.\ Rev.\ D {\bf 86} (2012) 073012
  [arXiv:1205.4018 [hep-ph]].

\bibitem{Fogli:2012ua}
  G.~L.~Fogli, E.~Lisi, A.~Marrone, D.~Montanino, A.~Palazzo and A.~M.~Rotunno,
  Phys.\ Rev.\ D {\bf 86} (2012) 013012
  [arXiv:1205.5254 [hep-ph]].

\bibitem{GonzalezGarcia:2012sz}
  M.~C.~Gonzalez-Garcia, M.~Maltoni, J.~Salvado and T.~Schwetz,
  arXiv:1209.3023 [hep-ph].

\bibitem {Itow}
\fussy
Y. Itow, talk at {\em 25th International Conference on Neutrino Physics
and Astrophysics} (Neutrino 2012), Kyoto, Japan, June 3--9, 2012.


\bibitem{Ayres:2004js}
  D.~S.~Ayres {\it et al.}  [NOvA Collaboration],
  hep-ex/0503053.

\bibitem{Itow:2001ee}
  Y.~Itow {\it et al.}  [T2K Collaboration],
  hep-ex/0106019.

\bibitem{Abe:2011ts}
  K.~Abe, T.~Abe, H.~Aihara, Y.~Fukuda, Y.~Hayato, K.~Huang, 
A.~K.~Ichikawa and M.~Ikeda {\it et al.},
  arXiv:1109.3262 [hep-ex].

\bibitem{Badertscher:2008bp}
  A.~Badertscher, T.~Hasegawa, T.~Kobayashi, A.~Marchionni, 
A.~Meregaglia, T.~Maruyama, K.~Nishikawa and A.~Rubbia,
  arXiv:0804.2111 [hep-ph].

\bibitem{Goon:2012if}
  J.~Goon {\it et al.}  [LBNE Collaboration],
  arXiv:1204.2295 [physics.ins-det].

\bibitem{LBNO}
  A.~Stahl, C.~Wiebusch, A.~M.~Guler, M.~Kamiscioglu, R.~Sever, 
A.~U.~Yilmazer, C.~Gunes and D.~Yilmaz {\it et al.},
  CERN-SPSC-2012-021.

\bibitem{Geer:1997iz}
  S.~Geer,
  Phys.\ Rev.\ D {\bf 57} (1998) 6989
   [Erratum-ibid.\ D {\bf 59} (1999) 039903]
  [hep-ph/9712290].

\bibitem{Bandyopadhyay:2007kx}
  A.~Bandyopadhyay {\it et al.}  [ISS Physics Working Group Collaboration],
  Rept.\ Prog.\ Phys.\  {\bf 72} (2009) 106201
  [arXiv:0710.4947 [hep-ph]].

\bibitem{Zucchelli:2002sa}
  P.~Zucchelli,
  Phys.\ Lett.\ B {\bf 532} (2002) 166.

\bibitem{Barger:2001yr}
  V.~Barger, D.~Marfatia and K.~Whisnant,
  Phys.\ Rev.\ D {\bf 65} (2002) 073023
  [hep-ph/0112119].

\bibitem{BurguetCastell:2001ez}
  J.~Burguet-Castell, M.~B.~Gavela, J.~J.~Gomez-Cadenas, P.~Hernandez and O.~Mena,
  Nucl.\ Phys.\ B {\bf 608} (2001) 301
  [hep-ph/0103258].

\bibitem{Minakata:2001qm}
  H.~Minakata and H.~Nunokawa,
  JHEP {\bf 0110} (2001) 001
  [hep-ph/0108085].

\bibitem{Fogli:1996pv}
  G.~L.~Fogli and E.~Lisi,
  Phys.\ Rev.\ D {\bf 54} (1996) 3667
  [hep-ph/9604415].

\bibitem{Yasuda:2004gu}
  O.~Yasuda,
  New J.\ Phys.\  {\bf 6} (2004) 83
  [hep-ph/0405005].

\bibitem{Bertolucci:2012fb}
  S.~Bertolucci, A.~Blondel, A.~Cervera, A.~Donini, M.~Dracos, 
D.~Duchesneau, F.~Dufour and R.~Edgecock {\it et al.},
  arXiv:1208.0512 [hep-ex].

\bibitem{MINSIS}
The MINSIS proposal,\\
{\tt http://www-off-axis.fnal.gov/MINSIS/index.html}.

\bibitem{Alonso:2010wu}
  R.~Alonso, S.~Antusch, M.~Blennow, P.~Coloma, A.~de Gouvea, 
E.~Fernandez-Martinez, B.~Gavela and C.~Gonzalez-Garcia {\it et al.},
  arXiv:1009.0476 [hep-ph].

\bibitem{Choubey:2011zzq}
  S.~Choubey {\it et al.}  [IDS-NF Collaboration],
  arXiv:1112.2853 [hep-ex].

\bibitem{Grossman:1995wx}
  Y.~Grossman,
  Phys.\ Lett.\  B {\bf 359} (1995) 141
  [arXiv:hep-ph/9507344].

\bibitem{GonzalezGarcia:2001mp}
  M.~C.~Gonzalez-Garcia, Y.~Grossman, A.~Gusso and Y.~Nir,
  Phys.\ Rev.\  D {\bf 64} (2001) 096006
  [arXiv:hep-ph/0105159].

\bibitem{Ota:2001pw}
  T.~Ota, J.~Sato and N.~a.~Yamashita,
  Phys.\ Rev.\  D {\bf 65} (2002) 093015
  [arXiv:hep-ph/0112329].

\bibitem{Davidson:2003ha}
  S.~Davidson, C.~Pena-Garay, N.~Rius and A.~Santamaria,
  JHEP {\bf 0303} (2003) 011
  [hep-ph/0302093].

\bibitem{Antusch:2006vwa}
  S.~Antusch, C.~Biggio, E.~Fernandez-Martinez, M.~B.~Gavela and J.~Lopez-Pavon,
  JHEP {\bf 0610} (2006) 084
  [arXiv:hep-ph/0607020].

\bibitem{FernandezMartinez:2007ms}
  E.~Fernandez-Martinez, M.~B.~Gavela, J.~Lopez-Pavon and O.~Yasuda,
  Phys.\ Lett.\  B {\bf 649} (2007) 427
  [arXiv:hep-ph/0703098].

\bibitem{Antusch:2009pm}
  S.~Antusch, M.~Blennow, E.~Fernandez-Martinez and J.~Lopez-Pavon,
  Phys.\ Rev.\  D {\bf 80}, 033002 (2009)
  [arXiv:0903.3986 [hep-ph]].

\bibitem{Aguilar:2001ty}
  A.~Aguilar {\it et al.}  [LSND Collaboration],
  Phys.\ Rev.\  D {\bf 64} (2001) 112007
  [arXiv:hep-ex/0104049].

\bibitem{AguilarArevalo:2007it}
  A.~A.~Aguilar-Arevalo {\it et al.}  [The MiniBooNE Collaboration],
  Phys.\ Rev.\ Lett.\  {\bf 98} (2007) 231801
  [arXiv:0704.1500 [hep-ex]].

\bibitem{AguilarArevalo:2010wv}
  A.~A.~Aguilar-Arevalo {\it et al.}  [The MiniBooNE Collaboration],
  Phys.\ Rev.\ Lett.\  {\bf 105} (2010) 181801
  [arXiv:1007.1150 [hep-ex]].

\bibitem{Mueller:2011nm}
  T.~A.~Mueller {\it et al.},
  Phys.\ Rev.\  C {\bf 83} (2011) 054615
  [arXiv:1101.2663 [hep-ex]].

\bibitem{Huber:2011wv}
  P.~Huber,
  arXiv:1106.0687 [hep-ph].

\bibitem{Mention:2011rk}
  G.~Mention, M.~Fechner, T.~Lasserre, T.~A.~Mueller, D.~Lhuillier, M.~Cribier and A.~Letourneau,
  Phys.\ Rev.\  D {\bf 83} (2011) 073006
  [arXiv:1101.2755 [hep-ex]].

\bibitem{Giunti:2007xv}
  C.~Giunti and M.~Laveder,
  Phys.\ Rev.\ D {\bf 77} (2008) 093002
  [arXiv:0707.4593 [hep-ph]].

\bibitem{Donini:2008wz}
  A.~Donini, K.~-i.~Fuki, J.~Lopez-Pavon, D.~Meloni and O.~Yasuda,
  JHEP {\bf 0908} (2009) 041
  [arXiv:0812.3703 [hep-ph]].

\bibitem{Kyberd:2012iz}
  P.~Kyberd {\it et al.}  [nuSTORM Collaboration],
  arXiv:1206.0294 [hep-ex].

\end{thebibliography}
\end{document}